\documentclass[reprint,showpacs,twocolumn,superscriptaddress,aps,prx]{revtex4-2}

\usepackage[latin1]{inputenc}
\usepackage{bbm}
\usepackage{bm}
\usepackage[usenames]{color}
\usepackage{multirow}
\usepackage{amssymb}
\usepackage{amsbsy}
\usepackage{mathtools}
\usepackage{amsmath}
\usepackage{stmaryrd}
\usepackage{graphicx}
\usepackage{soul}
\usepackage{epsfig}
\usepackage{placeins}
\usepackage{bbold}
\usepackage{braket}
\usepackage{blindtext}
\usepackage[colorlinks,linkcolor=blue,citecolor=blue,urlcolor=blue]{hyperref}

\usepackage{scalerel}
\usepackage{tikz}
\usetikzlibrary{calc}
\usetikzlibrary{patterns}
\usetikzlibrary{svg.path}
\definecolor{orcidlogocol}{HTML}{A6CE39}
\tikzset{
  orcidlogo/.pic={
    \fill[orcidlogocol] svg{M256,128c0,70.7-57.3,128-128,128C57.3,256,0,198.7,0,128C0,57.3,57.3,0,128,0C198.7,0,256,57.3,256,128z};
    \fill[white] svg{M86.3,186.2H70.9V79.1h15.4v48.4V186.2z}
                 svg{M108.9,79.1h41.6c39.6,0,57,28.3,57,53.6c0,27.5-21.5,53.6-56.8,53.6h-41.8V79.1z M124.3,172.4h24.5c34.9,0,42.9-26.5,42.9-39.7c0-21.5-13.7-39.7-43.7-39.7h-23.7V172.4z}
                 svg{M88.7,56.8c0,5.5-4.5,10.1-10.1,10.1c-5.6,0-10.1-4.6-10.1-10.1c0-5.6,4.5-10.1,10.1-10.1C84.2,46.7,88.7,51.3,88.7,56.8z};
  }
}

\newcommand\orcid[1]{\!%
  \href{https://orcid.org/#1}{%
    \mbox{%
      \scaleto{%
        \begin{tikzpicture}[yscale=-1,transform shape]
          \pic{orcidlogo};
        \end{tikzpicture}
      }{8pt}%
    }%
  }%
}

\begin{document}

\title{
Evidence for simple ``arrow of time functions" in closed chaotic quantum systems 
}
\author{Merlin F\"{u}llgraf~\orcid{0009-0000-0409-5172}
}
\email{mfuellgraf@uos.de}
\author{Jiaozi Wang
\orcid{0000-0001-6308-1950}
}
\email{jiaowang@uos.de}
\author{Jochen Gemmer
\orcid{0000-0002-4264-8548}
}%
\email{jgemmer@uos.de}
\affiliation{University of Osnabr\"{u}ck, Department of Mathematics/Computer Science/Physics, D-49076 Osnabr\"{u}ck, Germany}%
\date{\today}
\begin{abstract}

    Through an explicit construction, we assign to any infinite temperature autocorrelation  function $C(t)$  a set of functions $\alpha^n(t)$. The construction of  $\alpha^n(t)$ from $C(t)$ requires the first $2n$ temporal derivatives of $C(t)$ at times $0$ and $t$.
    Our focus is on  $\alpha^n(t)$ that  (almost) monotonously decrease; we call these ``arrows of time functions" (AOTFs). For autocorrelation functions of few-body observables we numerically observe the following: An AOTF  featuring a low $n$ may always be found unless the system is in or close to a nonchaotic regime with respect to a variation of some system  parameter. All $\alpha^n(t)$ put upper bounds to the respective autocorrelation functions, i.e.\ $\alpha^n(t) \geq C^2(t)$. Thus the implication of the existence of an AOTF  is comparable to that of the H-Theorem, as it indicates a directed approach to equilibrium.
    We furthermore argue that our numerical finding may to some extent be traced back to the operator growth hypothesis. This argument is laid out in the framework of the so-called recursion method.

\end{abstract}
\maketitle
\section{Introduction}

Is there a tendency in closed, nondriven quantum systems to approach equilibrium in a monotonous fashion? And if so, how would one even pinpoint all the somewhat vague notions in this question? Obviously this question is related to the so-called ``arrow of time". A full-fledged overview over all approaches to this subject, even restricted to the quantum realm, is clearly beyond the possibilities of the paper at hand. Thus, below we only very briefly introduce some concepts that appear to be in the vicinity or form the background  of  what shall be mainly presented later in the present paper. Readers well acquainted with these subjects may just skip this part.

\textit{\textcolor{black}{Arrow of time in closed quantum systems:}} While the full von Neumann entropy in closed system can never approach equilibrium (i.e. increase), other quantities like expectation values of certain observables, reduced density matrices, etc. may do so, in the sense that large deviations from  equilibrium become extremely rare, as captured by the term ``equilibration on average". Under rather mild conditions such a scenario is even likely to emerge, as the milestone concepts of typicality and the eigenstate thermalization hypothesis suggest \cite{popescu06,rigol08,gogolin16}. The caveat of ``equilibration on average"  with respect to an arrow of time is, that it only  establishes the low relative frequency of deviations from  equilibrium, but no temporal order for these rare events emerges. The deviations  could e.g. persist for a very long period in the beginning if they subsequently vanish for an even larger period, or (re)appear after a very long period of (apparent) closeness to equilibrium. For an illustration of the latter see Ref.\ \cite{knipschild20}, for an attempt to deal with the former see Ref.\ \cite{gpintos17}.

\textit{\textcolor{black}{Arrow of time in measured quantum systems:}} The von Neumann entropy is nondecreasing under projective measurements (if the measurement result is not read out) \cite{vNeumann96}. Since it is invariant under unitary evolution, it cannot decrease in a sequence of intermittent unitary evolutions and measurements. A continuous version of this scenario is often  captured by means of an open system description with a dephasing model \cite{breuer2007}. In such descriptions  the Kullback-Leibler divergence of the actual state and a stationary state is nonincreasing\ \cite{landi21}. Hence, for repeatedly measured systems this Kullback-Leibler divergence may be considered as a nonincreasing distance  to equilibrium and thus constitutes an arrow of time. However, the paradigm that only the external measurement drives systems to equilibrium is at odds with a number of recent experiments (e.g., on cold atoms see \cite{ueda20}) and the above concept of equilibration in closed quantum systems. 

\textit{\textcolor{black}{Arrow of time, observational entropy, Boltzmann entropy  and  consistent histories:}} This approach  may be regarded as a synthesis of the open and the closed system concepts. 
It is based on the observation that some dynamics of a (typically more complex) closed system  may appear as if the latter was subject to external measurements. In this case so-called ``consistent histories" emerge. Whether or not this is the case could in principle be answered by computing a so-called ``decoherence functional" \cite{dowker1992}, in practice this may be hard. However, if consistent histories emerge, the results from the open quantum system carry over and a suitably defined analog of the von Neumann entropy is nondecreasing \cite{Woo2000}. It has also been pointed out that in this case an even more accessible quantity called ``observational entropy" \cite{strasberg21,safranek21,strasberg23} is nondecreasing.  It has also been shown that for an ideal quantum gas the Boltzmann entropy, a quantity conceptually similar to observational entropy, is nondecreasing \cite{pandey23}.
These approaches are closest to the main subject of the  present paper, as they target monotonous\\ (entropy-)functions  that are entirely defined within closed, non-driven systems. However, while the ``arrow of time functions" (which will be  defined below) are no thermodynamic state functions,  their monotonicity appears to prevail in ``pure quantum settings" where no consistent histories emerge. The latter does not hold for observational entropy  \cite{strasberg23}.

In the paper at hand we are working within the closed-system framework. However, we aim at going beyond the common lore of  ``scarceness of excursions from equilibrium"  towards establishing a time-local arrow of time in the spirit of the Boltzmann H-theorem\ \cite{Boltzmann1970}.  While a substantial part of this effort consists in numerical observations,  we attempt to trace these finding partially back to the operator growth hypothesis\ \cite{parker19} in the second part of the present paper. Throughout this analysis we take the ``chaoticity" of  a system into account which we measure by the gap ratio \cite{oganeysan07}.

As we find that specifically simple H-theorem type functions emerge (primarily, not exclusively)  for chaotic systems, the central quantities developed below add to the limited number of dynamical manifestations of chaos, like, e.g., the correlation hole, described in  the literature\ \cite{leviandier86,prosen02,torres17,schmitt18}. 

\textcolor{black}{(\textit{Some) other approaches to irreversibility in closed quantum systems:}} Several considerations suggest  that autocorrelation functions in chaotic (quantum) systems are well expandable in terms of (a few) exponentials with negative real parts in the exponents, either from the beginning, or at least in the long run \cite{prosen02-2,Mori24,teretenkov25,dodelson24}. This concept goes under the name of Ruelle-Pollicott resonances, pseudomodes or quasinormal modes. If such an approach applies a direction of time is naturally established. 

Another line of research is based on the so-called volume or Hertz entropy. The latter features various thermodynamically desirable properties. There is also evidence that it is nondecreasing in certain scenarios \cite{CAMPISI2008181,campisi08,joshi13,Campisi16}.

The paper is organised as follows. In Sec. \ref{sec:construction} we introduce our construction for arrow of time functions together with the notion of \textit{intricacy} as a central quantity in the paper at hand. In Sec.\ \ref{sec:emergence} we present numerical data for several pertinent quantum systems and observables. Translating our construction into the setting of the recursion method in Sec.\ \ref{sec:lanczos} we introduce a replacement model to link our findings to the so-called operator growth hypothesis in Sec.\ \ref{sec:structures}. We conclude by summarizing our work in Sec.\ \ref{sec:conclusion}.

\section{Construction of arrow of time functions\label{sec:construction}}

With the goal to capture the emergent irreversibility of the equilibration process of an observable $\mathcal{O}$ and therefore indicating the direction of time we suggest in the following a scheme to assign the autocorrelation $C(t)=\frac{1}{d}\text{tr}\{\mathcal{O}(t)\mathcal{O}\}$ a function describing an effective deviation to equilibrium. Here $d$ is the dimension of the respective Hilbert space. 
The scheme iteratively computes a set of $M$ functions $\{\varphi_j(t)\}_{j\ge0}$ starting from $C(t)$ as its germ. Starting with $C(t)=:\varphi_0(t)$ the subsequent $\varphi_j$ follow via 
\begin{align}
    \begin{split}
        \textcolor{black}{\tilde{\varphi}_j(t)}&\textcolor{black}{:=-\partial_t \varphi_{j-1}(t)+b_{j-1} \varphi_{j-2}(t)},\\
        \textcolor{black}{b_j^{2}}&\textcolor{black}{:=\langle\tilde{\varphi}_j(t), \tilde{\varphi}_j(t)\rangle,}\\
        \textcolor{black}{\varphi_{j}(t)}&\textcolor{black}{:=b_j^{-1}\tilde{\varphi}_j(t),}
    \end{split}\label{construction}
\end{align}
\textcolor{black}{with $b_0=0$. The inner product $\langle -,-\rangle$ is defined on the basis $\{\partial_t^k C(t)\}$ via $\langle \partial_t^n C(t),\partial_t^m C(t)\rangle:=(-1)^n\partial_t^{n+m}C(t)\vert_{t=0}$ and constitutes a well-defined scalar product on the space spanned by the first $M$ derivatives of    $C(t)$.} \textcolor{black}{For further details, see App.\ \ref{app:details}.} With these $\{\varphi_j(t)\}$ we construct
\begin{align}
    \alpha^n(t):=\sum_{j=0}^n \varphi_{j}^2(t).\label{aot}
\end{align}
We then determine $\mathcal{I}$ as the smallest value for which $\alpha^\mathcal{I}$ becomes monotonously decreasing up to an accumulated error \textcolor{black}{ $\delta$  until the relaxation time $\tau_\mathcal{I}$} \footnote{\textcolor{black}{We infer the relaxation time $\tau_{\mathcal{I}}$ as the time after which the function $\alpha^n(t)$ has permanently fallen below some relaxation threshold. Concretely we determine $\tau_\mathcal{I}$ as the earliest time such that $\alpha^n(t)\le 1/200$ $\forall\ t\ge \tau_{\mathcal{I}}$}}. \textcolor{black}{The "accumulated error" of a function $\alpha^n(t)$ therefore refers to the sum of all ascends of the function until $\tau_{\mathcal{I}}$.} (For the issue of Poincar\'{e} recurrences, cf. Ref. \footnote{\textcolor{black}{Strictly speaking such a time $\tau_\mathcal{I}$ does not exist for any finite system, since there will always be Poincar\'{e} recurrences.  These, however occur by magnitudes later than any time scale considered in the paper at hand.  
Technically $\tau_\mathcal{I}$ is always substantially smaller than the largest time we can access numerically.}}.) In order to practically determine $\mathcal{I}$ for some $C(t)$ we use numerical methods and simply check the monotonicity of $\alpha^n(t)$ while increasing $n$ ``one by one", see also Fig.\ \ref{fig:panel_wheel} (\textit{lower-middle} row). The value of $\delta$ should be suitably chosen in a sense that is supposed to become clear in the remainder of this paper. For the entirety of the paper at
hand we set $\delta= 1/200$. We refer to the quantity $\mathcal{I}$ as  \textit{intricacy} for reasons which will also become clear below. In passing we mention that the $\alpha^n(t)$ have an intrinsic connection to the ``Krylov-complexity" \cite{parker19} as detailed in Sect. \ref{subsec:currents}.

For given $C(t), \delta$ a finite  $\mathcal{I}$ may or may not exist. We interpret $\alpha^\mathcal{I}(t)$ as a \textit{deviation-from-equilibrium}, e.g., since it limits possible values of $C(t)$ through $\alpha^\mathcal{I}(t) \geq C^2(t)$. Due to its almost monotonous decrease, we also refer to $\alpha^\mathcal{I}(t)$ as an  \textit{arrow of time function} (AOTF).  The  degree to which the function $\alpha^\mathcal{I}(t)$ may be described as  time-local depends on $\mathcal{I}$. From the construction one readily infers that $2\mathcal{I}$ temporal derivatives at $t=0$ and $\mathcal{I}$ temporal derivatives at $t$ are required to actually compute $\alpha^\mathcal{I}(t)$. Thus, for not too large $\mathcal{I}$, we refer to $\alpha^\mathcal{I}(t)$ as time-local with respect to $C(t)$. Physically numerous but in this context somewhat trivial cases emerge for autocorrelation functions $C(t)$ which are themselves monotonously decaying, regardless of the decay time. In this case $\alpha^0(t)$ is an AOTF. While this is by no means excluded, this investigation focuses on more complex $C(t)$ that do not decay monotonously but induce AOTFs with finite $\mathcal{I}$.

To contextualize AOTFs we briefly comment on their comparability to observational entropy \cite{strasberg21,safranek21} as both objects are closely related to irreversibility and defined within closed quantum systems. Let $C(t)$  correspond to the (infinite temperature) autocorrelation function of some observable $\mathcal{O}$. Consider an initial state $\rho \propto \exp(\beta \mathcal{O})$ with small $\beta$. Then $\alpha^\mathcal{I}(t)$ may be computed from the expectation values $\langle \partial_t^n \mathcal{O}(t)\rangle$ with $0\leq n \leq \mathcal{I}$. Thus, just like observational entropy is based on the knowledge of a case dependent set of expectation values, so is $\alpha^\mathcal{I}(t)$ in this case. Note, however, that the $\langle \partial_t^n \mathcal{O}(t)\rangle$ must not be interpreted as probabilities, as neither $\langle \partial_t^n \mathcal{O}(t)\rangle > 0$ nor $\sum_n\langle \partial_t^n \mathcal{O}(t)\rangle =1$ holds. Furthermore, unlike (observational) entropies, the  $\alpha^\mathcal{I}(t)$ are no thermodynamic state functions, nevertheless they indicate the direction of time in the dynamics of  $C(t)$ (or, in the above special case $\langle \mathcal{O}(t)\rangle$) including the prediction of  an eventual stationary value. Thus, the concept of the AOTFs  presented here should be considered as complementary to the concept of observational entropy. Other than for observational entropy, the monotonicity of  the  $\alpha^\mathcal{I}(t)$  (with low $\mathcal{I}$) appears not to depend on slowness of the observable or the emergence of consistent histories, cf. Ref.\ \cite {strasberg23, gemmer14} as will become evident below.

\section{Numerical evidence for arrows of time\label{sec:emergence}}
In this paper we investigate autocorrelation functions of few-body observables $\mathcal{O}$ in several quantum systems $\mathcal{H}(\sigma)$ which exhibit deviations from the fully chaotic regime moderated by specific system parameters $\sigma$. We quantify the notion of chaoticity by means of the mean gap ratio $\langle r\rangle$ of adjacent level spacings as a standard indicator for quantum chaos \cite{oganeysan07}. Concretely by
\begin{align}
    \langle r\rangle=\frac{1}{d}\sum_{k}\frac{\min(s_k,s_{k+1})}{\max(s_k,s_{k+1})},
\end{align}
where $s_k=\epsilon_{k+1}-\epsilon_{k}$ with the $\epsilon_k$ as the eigenenergies of the considered system. 
Exploiting respective symmetries, we truncate the system to a specific symmetry-separated subspace.
For fully chaotic systems, the mean gap ratio is numerically determined as $\langle r\rangle_{\text{GOE}}\approx0.53$ by means of large random matrices from a Gaussian orthogonal ensemble (GOE). Conversely, for integrable systems exhibiting a Poissonian spectrum, $\langle r\rangle_P\approx0.386$ \cite{oganeysan07}.

While tracking the respective gap ratio along $\sigma$, we also compute the autocorrelation $C(t)$ for different system parameters. For these autocorrelation functions we determine $\mathcal{I}$ as the intricacy of the arrow of time function as described in Sec.\ \ref{sec:construction}.

For details on the numerical method applied in this work, we refer to Appendix\ \ref{numerical_method}.
\subsection{Models and observables\label{subsec:models}}
\textit{Two-chain model:}
As a first model we consider two (mixed-field) Ising chains with local coupling of the respective $\sigma^z$-terms moderated by the interaction strength $\lambda$ and with periodic boundary conditions. \textcolor{black}{For the entirety of this paper we set $\hbar=1$.} The Hamiltonian reads
\begin{align}
    \mathcal{H}(\lambda)&=\mathcal{H}_1+\lambda\mathcal{H}_I+\mathcal{H}_2,\label{ladder}
\end{align}
with the Ising Hamiltonians
\begin{align}
    \mathcal{H}_a=\sum_k h_{a,k},
\end{align}
with $a=1,2$, local constituents
\begin{align}
\begin{split}
    h_{a,k}&=J\sigma_{a,k}^{z}\sigma_{a,k+1}^{z}+\frac{B_x}{2}\left( \sigma_{a,k}^x+\sigma_{a,k+1}^x\right)\\&+\frac{B_z}{2}\left(\sigma_{a,k}^z+\sigma_{a,k+1}^z\right),\label{localham}
\end{split}
\end{align}
and the interaction
\begin{align}
    \mathcal{H}_I=\sum_{k} \sigma_{1,k}^z \sigma_{2,k}^z.
\end{align}
We fix the parameters of the model as $(J,B_x,B_z)=(1,1,0.5)$ and vary the interaction strength $\lambda$. For the computation we consider a total size $L=18$. 

\begin{figure}[t]
	 \includegraphics[width=0.35\textwidth]{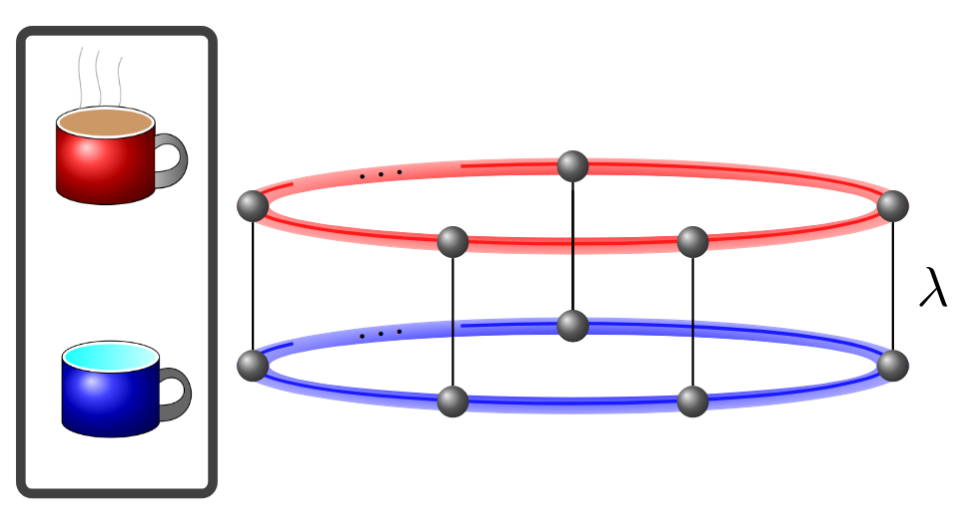}
 \caption{\label{fig_mugs}Sketch of the archetypal example of two macroscopic bodies in contact (here: two cups of coffee) described by the Hamiltonian given by (\ref{ladder}).}
 \end{figure}

\begin{figure}[ht]
	 \includegraphics[width=0.5\textwidth]{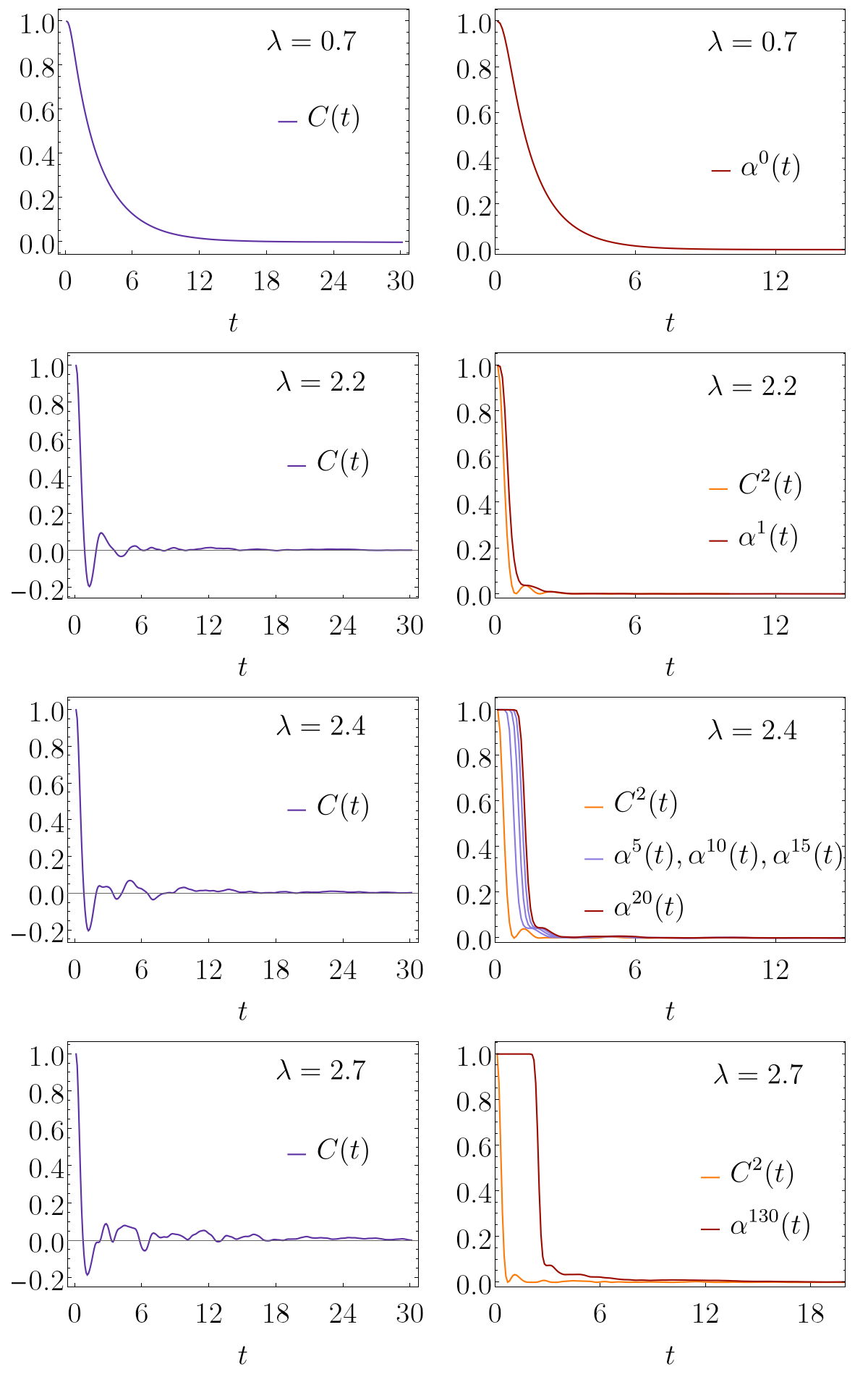}
 \caption{\label{fig:panel_wheel}Autocorrelation functions $C(t)$ (\textit{left}) and their respective squares $C^2(t)$ and AOTFs  $\alpha^\mathcal{I}(t)$ (\textit{right}) for the energy imbalance in the two-chain-model (\ref{ladder}) for different interaction strengths $\lambda$. \textit{Top:} At $\lambda=0.7$ (chaotic regime, see Fig.\ \ref{rR_wheel})  $C(t)$ itself is monotonously decreasing and hence \ $\alpha^0(t)$ is an  AOTF . \textit{Upper-middle:} At $\lambda=2.2$ (chaotic regime, see Fig.\ \ref{rR_wheel}) $C(t)$ no longer  decreases monotonously. However,   $\alpha^1(t)$ does, thus two derivatives w.r.t. time are required to construct an AOTF. \textit{Lower-middle:} At interaction strength $\lambda=2.4$ (nonchaotic regime, see Fig.\ \ref{rR_wheel}) the intricacy $\mathcal{I}$ is 20. We also depict some intermediate functions $\alpha^n(t)$, (here only $n=5,10,15$ to avoid cluttering) that fall short of the required monotonicity standard explained in Sec.\ \ref{construction}. \textit{Bottom:} In the nonchaotic regime at $\lambda=2.7$ (see Fig.\ \ref{rR_wheel}) a total of 260 derivatives w.r.t.\ time are required to construct the AOTF $\alpha^{130}(t)$.}
 \end{figure}

\begin{figure}[h]
	 \includegraphics[width=0.5\textwidth]{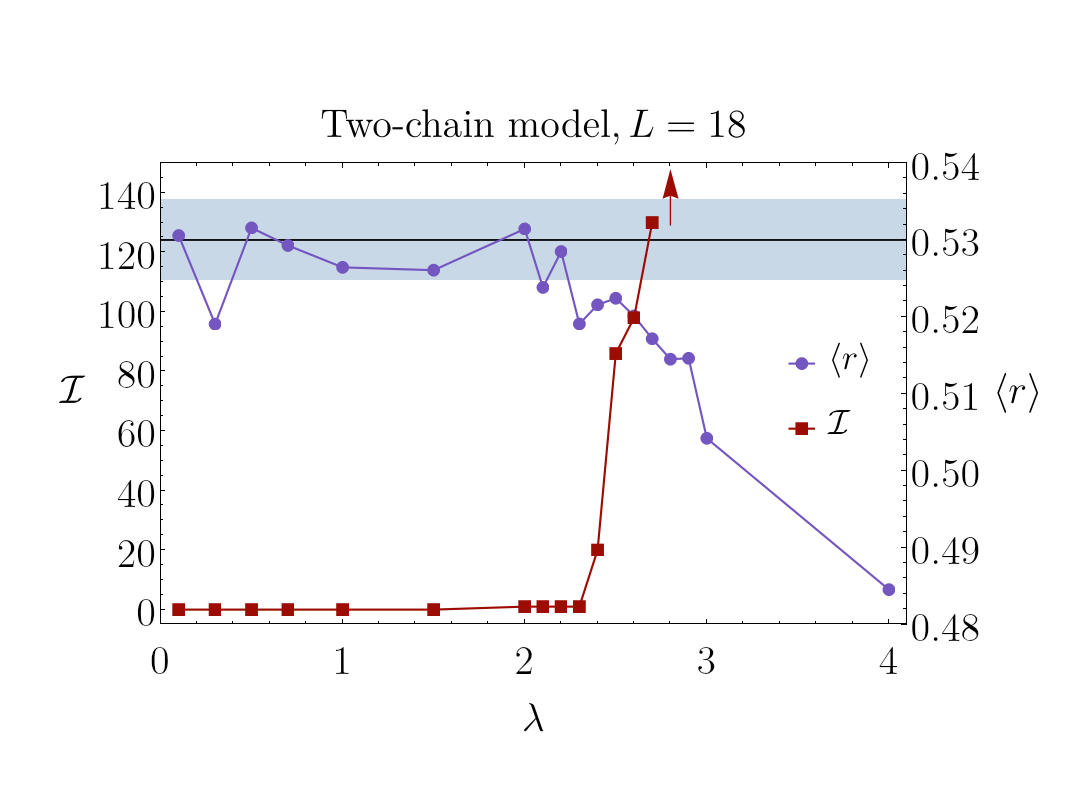}
  \caption{\label{rR_wheel}AOTF intricacy $\mathcal{I}$ and gap ratio $\langle r\rangle$ for different interaction strengths $\lambda$ in the site-wise coupled two-chain-model with $L=18$. The black line indicates the chaotic benchmark $\langle r\rangle_{\text{GOE}}=0.53$ and the shaded area $1\%$ deviation around it. The red arrow indicates that for larger $\lambda$ our numerical method suffices to determine resp. $\mathcal{I}$-values to exceed 200.}
\end{figure}

As the model mimics the pertinent scenario of the equilibration of two macroscopic bodies in contact (see Fig.\ \ref{fig_mugs}), we consequently consider the energy imbalance
\begin{align}
    \mathcal{O}=\mathcal{H}_1-\mathcal{H}_2. \label{balanceenergy}
\end{align}
For low $\lambda$ an  (almost) exponentially decaying $C(t)$ results, see Fig.\ \ref{fig:panel_wheel}.  This is reminiscent of the decay of  a temperature difference between two macroscopic bodies in mutual thermal contact, but insulated otherwise. The application of  a standard projection operator technique in the Markovian, weak-coupling  limit would find this result \cite{breuer2007}. The intricacy is $\mathcal{I}=0$.  However, as we increase $\lambda$ we leave the Markovian regime and at $\lambda \sim2$  $C(t)$ is no longer monotonously decaying but exhibits a more intricate form, see Fig.\ \ref{fig:panel_wheel}. Nevertheless the intricacy is moderate,  $\mathcal{I}=1$ and $\langle r \rangle $ is still close to the GOE value, cf. Fig.\ \ref{rR_wheel}.     (For an assessment of the observational entropy approach to this scenario, see Ref.\  \footnote{We expect the emergence of consistent histories and Markovianity to be absent at (and above) $\lambda \approx 2$, hence we do not expect a pertinent observational entropy to be monotonously decreasing.}.) If $\lambda$ is increased further, the gap ratio $\langle r \rangle $ starts to sink below the GOE value (since the model approaches a situation of decoupled dimers) and the intricacy rapidly  increases towards much higher values,
see Figs.\ \ref{fig:panel_wheel},\ref{rR_wheel}.

For this model we can rule out arrows of time for $\mathcal{I}\le200$  for $\lambda\gtrsim2.8$. Conversely, in the chaotic regime we find arrows of time with low intricacy giving rise to a \textit{time-local} description of the approach to equilibrium as only a low finite number of derivatives of the autocorrelation function $C(t)$ at times $0$ and $t$ are sufficient to construct an AOTF.

\begin{figure}[h]
	 \includegraphics[width=0.5\textwidth]{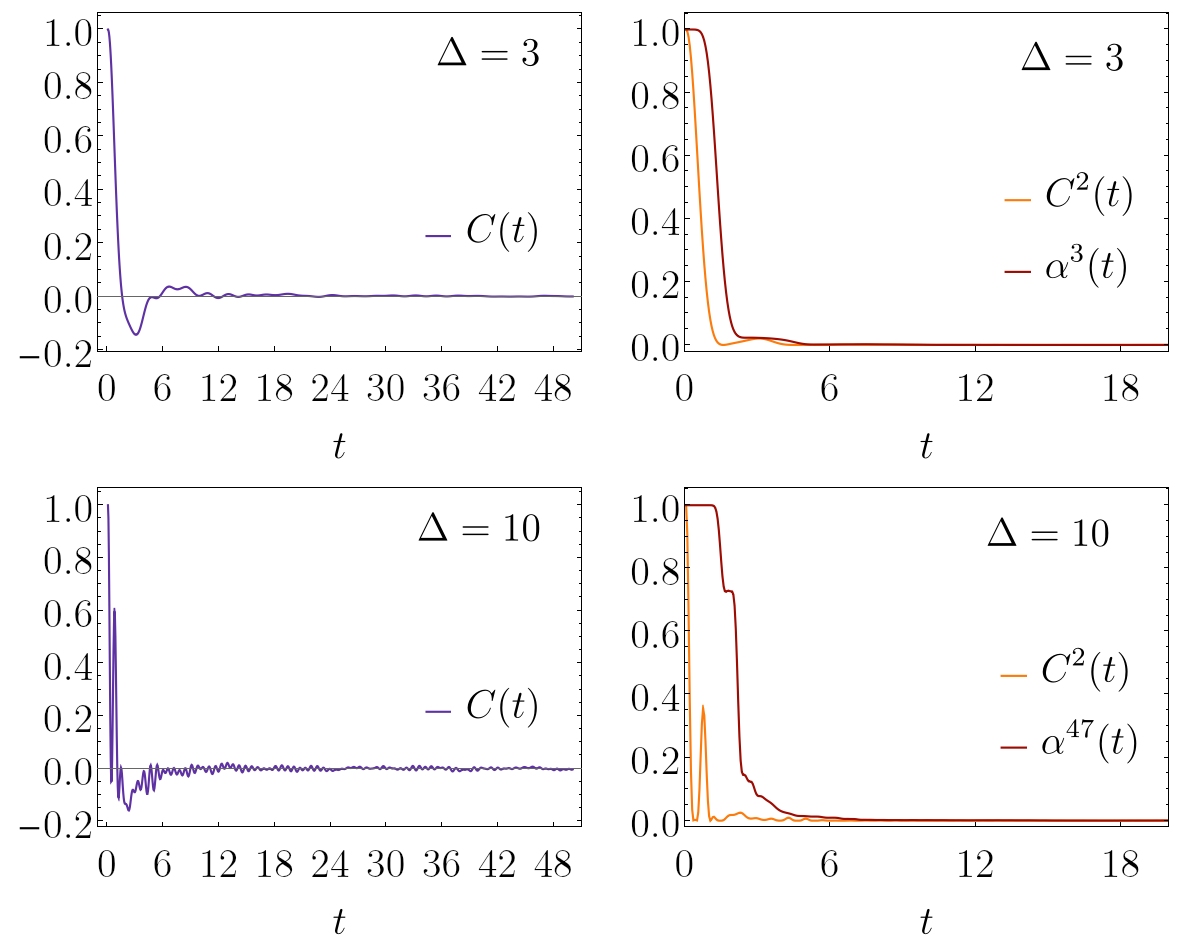}
 \caption{\label{fig:panel_xxz}Autocorrelation functions $C(t)$ (\textit{left}) and their respective squares $C^2(t)$ and AOTFs $\alpha^\mathcal{I}$ (\textit{right}) for the magnetisation current in the XXZ model (\ref{xxz-hamiltonian}) for different nearest-neighbour interactions $\Delta$. \textit{Top:} At $\Delta=3$  (chaotic regime, see Fig.\ \ref{rR_xxz}) $C(t)$ does not decay monotonously  but the AOTF $\alpha^3(t)$ does. \textit{Bottom:} Further away from the chaotic regime, at $\Delta=10$, an AOTF is reached only at $\mathcal{I}=47$, i.e. $\alpha^{47}(t)$ decays monotonously.} 
\end{figure}

\begin{figure}[h]
	 \includegraphics[width=0.5\textwidth]{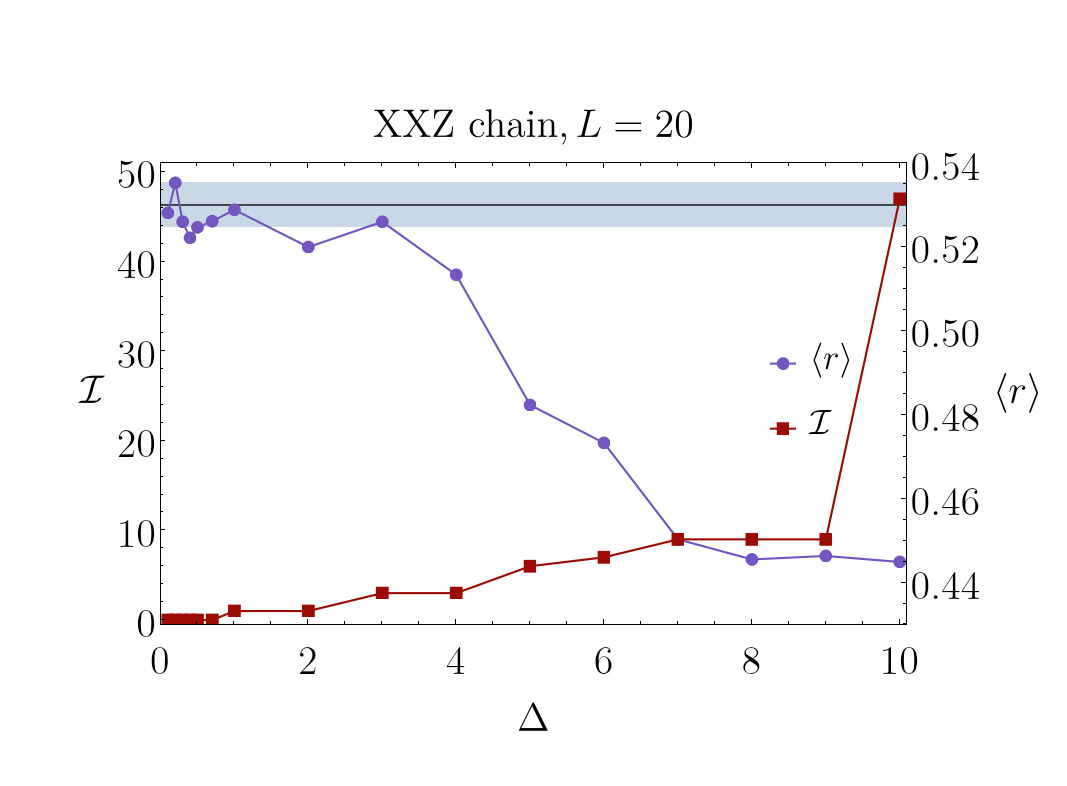}
  \caption{\label{rR_xxz}AOTF intricacy $\mathcal{I}$ and gap ratio $\langle r\rangle$ for different nearest-neighbour couplings $\Delta$ in the XXZ Heisenberg chain with $L=20$. The black line indicates the chaotic benchmark $\langle r\rangle_{\text{GOE}}=0.53$ and the shaded area $1\%$ deviation around it.}
\end{figure}

\textit{XXZ chain:}
Next, we consider an XXZ-type Heisenberg chain with length $L=20$ and periodic boundary conditions, modelled by the Hamiltonian
\begin{align}
    \mathcal{H}(\Delta)=J\sum_{k=1}^{L}s_k^x s_{k+1}^x+s_k^y s_{k+1}^y+\Delta s_k^z s_{k+1}^z+\Delta^\prime s_{k}^z s_{k+2}^z,\label{xxz-hamiltonian}
\end{align}
in which we set $(J,\Delta^\prime)=(1,0.5)$ and where $s_k^{a}=\sigma_k^{a}/2$ denote the spin-matrices. We vary the nearest-neighbour interaction term $\Delta$. Introducing the next-nearest interaction $\Delta^\prime$ breaks Bethe-Ansatz integrability \cite{bertini2021-1,takashi99}. However, for $\Delta\gg\Delta^\prime$ the system again approaches the integrable regime, see Fig.\ \ref{rR_xxz}. For any isotropies $\Delta,\Delta^\prime$ the total magnetization is conserved\ \cite{bertini2021-1}, motivating the study of the respective current 
\begin{align}
    \mathcal{O}=\mathcal{J}_S=\sum_k s^x_k s^y_{k+1}-s^y_k s^x_{k+1}
\end{align}
as an observable of interest. Depictions of respective AOTFs  are given in Fig.\ \ref{fig:panel_xxz}. Again, we find that for the chaotic system the autocorrelation functions allow  for simple AOTFs  with $\mathcal{I}\approx 5$. The intricacy $\mathcal{I}$ does grow to large values of $\mathcal{I}>40$ but this happens deeper in the nonchaotic regime, at $\Delta \approx 10$, see Fig.\ \ref{rR_xxz}.

\textit{Ising chain:}
As a third model we turn to a transverse Ising chain with length $L=18$ with a tilted magnetic field as very generic example for a chaotic quantum model. Its Hamiltonian is given by
\begin{align}
 \mathcal{H}(B_x)&=\sum_k h_k,\label{isingham}
\end{align}
with the local energies
\begin{align}
\begin{split}
        h_{k}&=J\sigma_{k}^{x}\sigma_{k+1}^{x}+\frac{B_z}{2}\left(\sigma_{k}^z+\sigma_{k+1}^z\right)+\frac{B_x}{2}\left(\sigma_{k}^x+\sigma_{k+1}^x\right).
\end{split}
\end{align}
We choose the parameter set as $(J,B_z)=(1,-1.05)$ and employ periodic boundary conditions. We vary $B_x$ breaking integrability as for $B_x=0$ the system is integrable whereas for $B_x\neq0$ the system enters the chaotic regime. We investigate the arrows of time of several kinds of physically interesting observables. First we consider
\begin{align}
    \mathcal{O}_{1}\propto\sum_k\cos\left(\pi k\right)h_k 
\end{align}
representing the fast-mode of the energy density-wave operator. This operator shows a quick relaxation. 

\begin{figure}[h]
	 \includegraphics[width=0.5\textwidth]{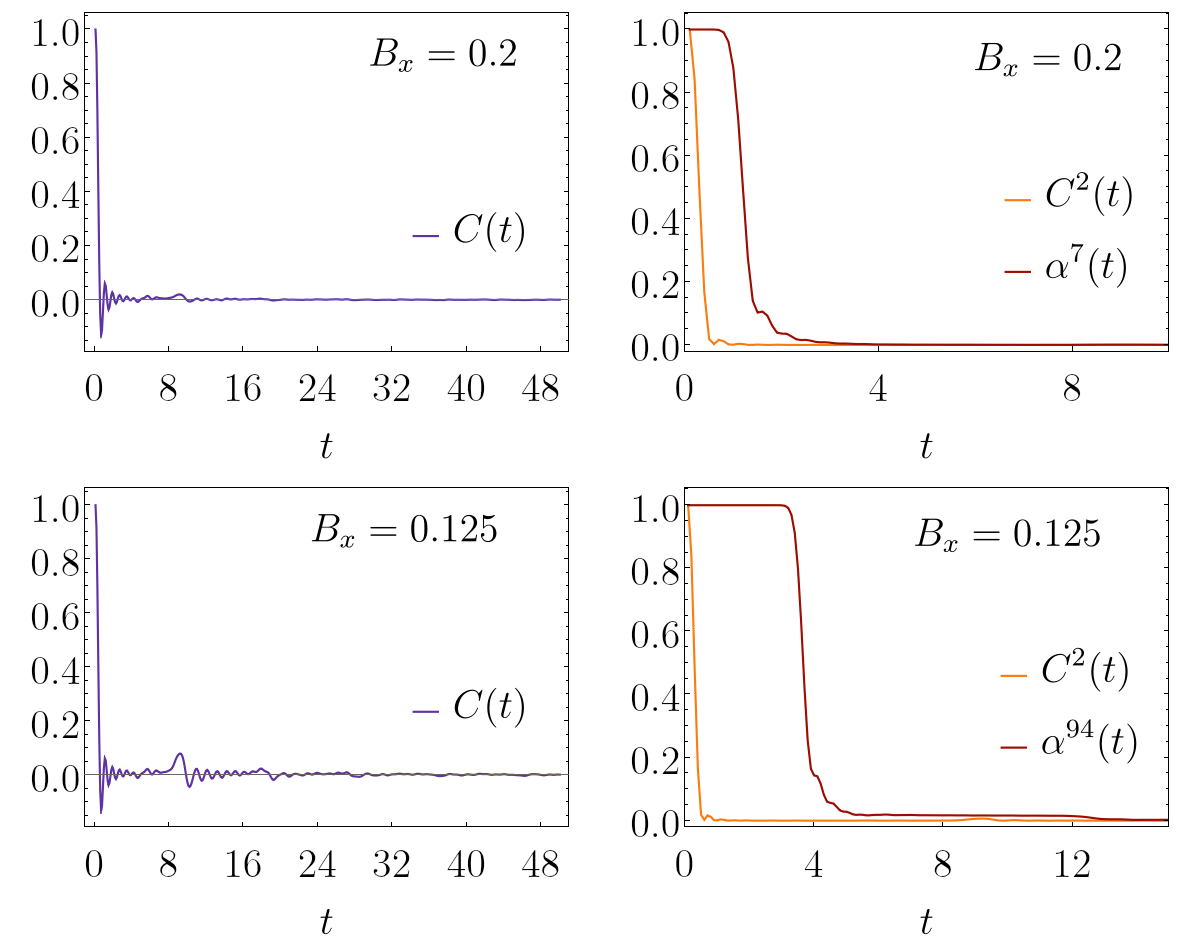}
 \caption{\label{fig:panel_ising}Autocorrelation functions $C(t)$ (\textit{left}) and their respective squares $C^2(t)$ and AOTFs $\alpha^\mathcal{I}$ (\textit{right}) for the orthogonal observable $\mathcal{O}_2$ in the tilted field Ising chain (\ref{isingham}) for different magnetic fields $B_x$. \textit{Top:} At $B_x=0.2$ (chaotic regime, see Fig.\ \ref{fig:rR_ising}) the autocorrelation function allows for an AOTF with intricacy $R=7$. \textit{Bottom:} Approaching the fully integrable regime at $B_x=0$ (see Fig.\ \ref{fig:rR_ising}), we find for $B_x=0.125$ an AOTF with $\mathcal{I}=94$.} 
\end{figure}

Moreover we also examine the 2-local observable
\begin{align}
    \mathcal{O}_2\propto\sum_k 1.05\ \sigma_k^x\sigma_{k+1}^x+\sigma_k^z.
\end{align}
In Fig.\ \ref{fig:panel_ising} we depicted exemplary autocorrelation functions of $\mathcal{O}_2$ as well as respective AOTFs  for particular magnetic fields $B_x$ in the chaotic regime as well as close to the integrable point.

Lastly, we include in our investigation the energy current
\begin{align}
   \mathcal{O}_3=\mathcal{J}_E=B_{x}\sum_{k}\sigma_{k}^{y}(\sigma_{k+1}^{z}-\sigma_{k-1}^{z}).
\end{align}
Before delving into the analysis of our findings for this model, some remarks are in order. In the integrable Ising chain with $B_x=0$ the current $\mathcal{J}_E$ is an integral of motion. Consequently the respective autocorrelation function never relaxes and no intricacy $\mathcal{I}$ can be inferred subject to the above scheme (\ref{construction}). At small magnetic fields $B_x$ a possible relaxation is beyond the time scale we consider, hence we do not include these points here.

\begin{figure}[h]
	 \includegraphics[width=0.5\textwidth]{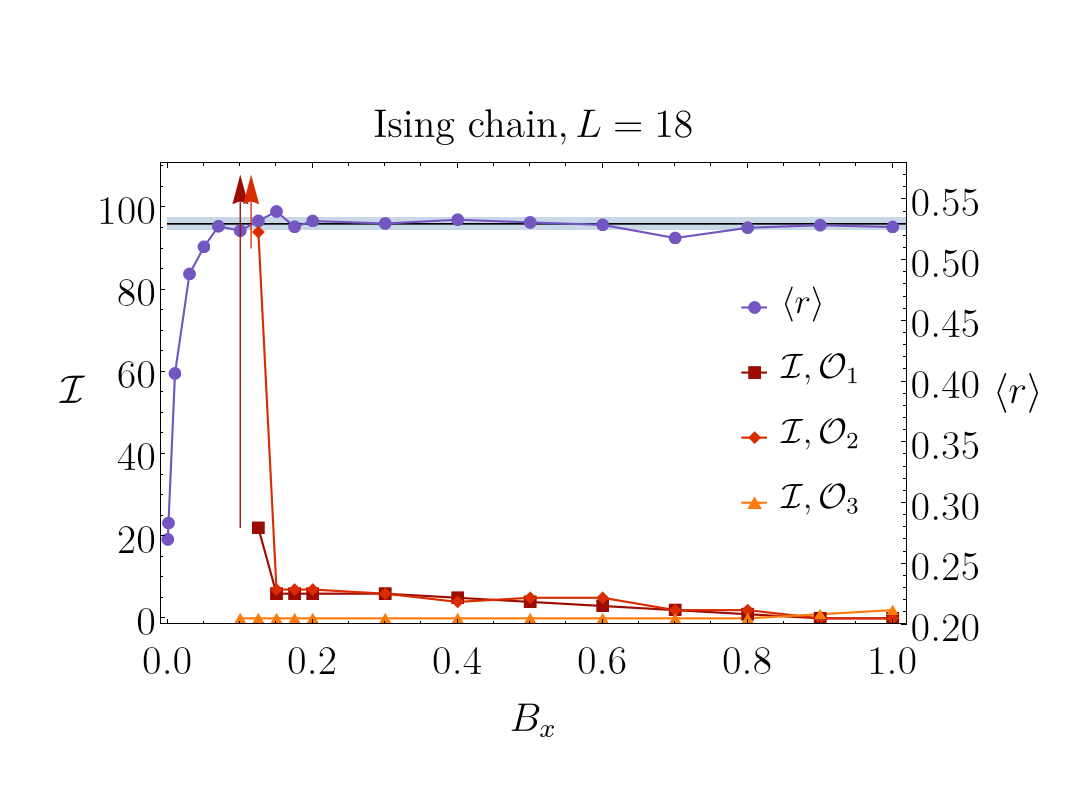}
  \caption{\label{fig:rR_ising}AOTF intricacy $\mathcal{I}$ and gap ratio $\langle r\rangle$ for magnetic fields $B_x$ in the tilted field Ising chain with $L=18$. The black line indicates the chaotic benchmark $\langle r\rangle_{\text{GOE}}=0.53$ and the shaded area $1\%$ deviation around it. The arrows indicate that for smaller $B_x$ our numerical method suffices to determine resp. $\mathcal{I}$-values to exceed 100.}
\end{figure}

Whereas for $B_x=0$ the model (\ref{isingham}) is integrable, introducing a non-zero tilted field $B_x$ breaks integrability and the system becomes chaotic, see Fig.\ \ref{fig:rR_ising}. Again, deep in the chaotic regime we find simple, low-intricacy AOTFs for all considered observables. Approaching the integrable point as $B_x\searrow0$, for the observables $\mathcal{O}_1$ and $\mathcal{O}_2$ the intricacies $\mathcal{I}$ grow, eventually exceeding the scope of our numerical analysis at $B_x\lesssim0.125$ indicating intricacies $\mathcal{I}>100$. 
For the energy current $\mathcal{O}_3$ the intricacies are low for every considered magnetic field $B_x$. 

\begin{figure}[h]
\includegraphics[width=0.5\textwidth]{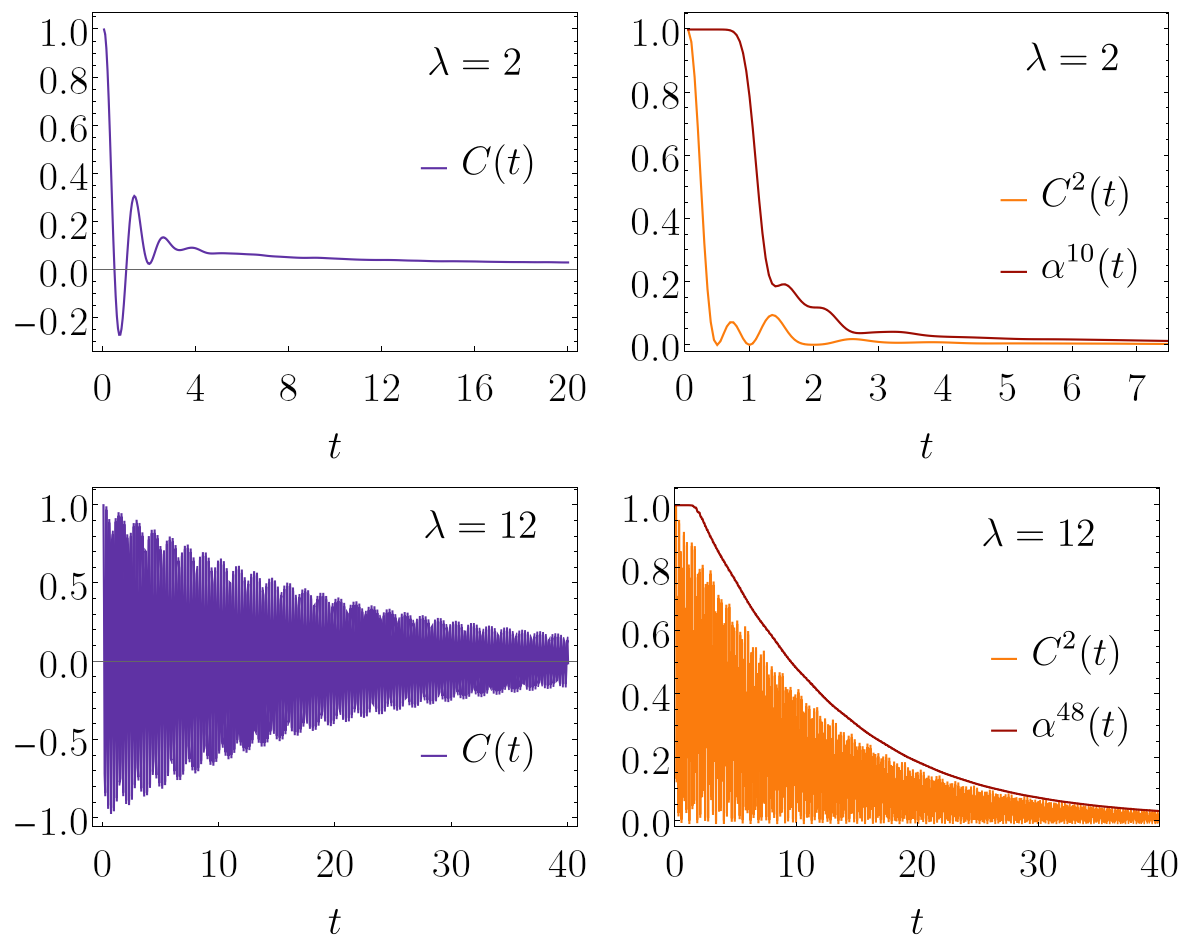}

  \caption{\label{fig:bath_sz}Autocorrelation functions $C(t)$ (\textit{left}) and their respective squares $C^2(t)$ and AOTFs $\alpha^\mathcal{I}$ (\textit{right}) for the observable \textcolor{black}{$\sigma_S^z$} in the spin bath model (\ref{spinbathhamilton}) for different couplings $\lambda$. \textit{Top:} At small coupling $\lambda$, here: $\lambda=2$ (chaotic regime, see Fig.\ \ref{fig:rR_spinbath}) we find an AOTF with intricacy $\mathcal{I}=10$.  \textit{Bottom:} Approaching the chaotic regime again at $\lambda = 12$, the intricacy of the respective AOTF drops to $\mathcal{I}=48$ again falling within our numerical scope.}
\end{figure}

\textit{Spin-bath model:}
Lastly we investigate the case of a single spin-1/2 coupled to an Ising spin bath of length $L=14$ with periodic boundary conditions. This system is described by the Hamiltonian
\begin{align}
\mathcal{H}(\lambda)=\mathcal{H}_S+\lambda\mathcal{H}_I+\mathcal{H}_B,\label{spinbathhamilton}
\end{align}
where $\lambda$ moderates the interaction of the spin and the bath via a coupling of the respective $x$-components. Concretely, we have
\begin{align}
    \mathcal{H}_S&=\sigma_S^z\\
    \mathcal{H}_I&=\sigma_S^x\otimes\sigma_1^x\\
    \mathcal{H}_B&= J\sum_j \sigma_j^z \sigma_{j+1}^z+B_x\sigma_j^x+B_z\sigma_j^z,
\end{align}
where we choose the parameters as $(\omega,J,B_x,B_z)=(1,1,1,0.5)$.
Varying the coupling strength $\lambda$ we compute the autocorrelation for the $x$ and $z-$component of coupled spin respectively, i.e.\ $\mathcal{O}^{(x)}=\sigma_S^x$ and $\mathcal{O}^{(z)}=\sigma_S^z$. See Fig.\ \ref{fig:bath_sz} for exemplary AOTFs of $\sigma_S^z.$

\begin{figure}[h]
	 \includegraphics[width=0.5\textwidth]{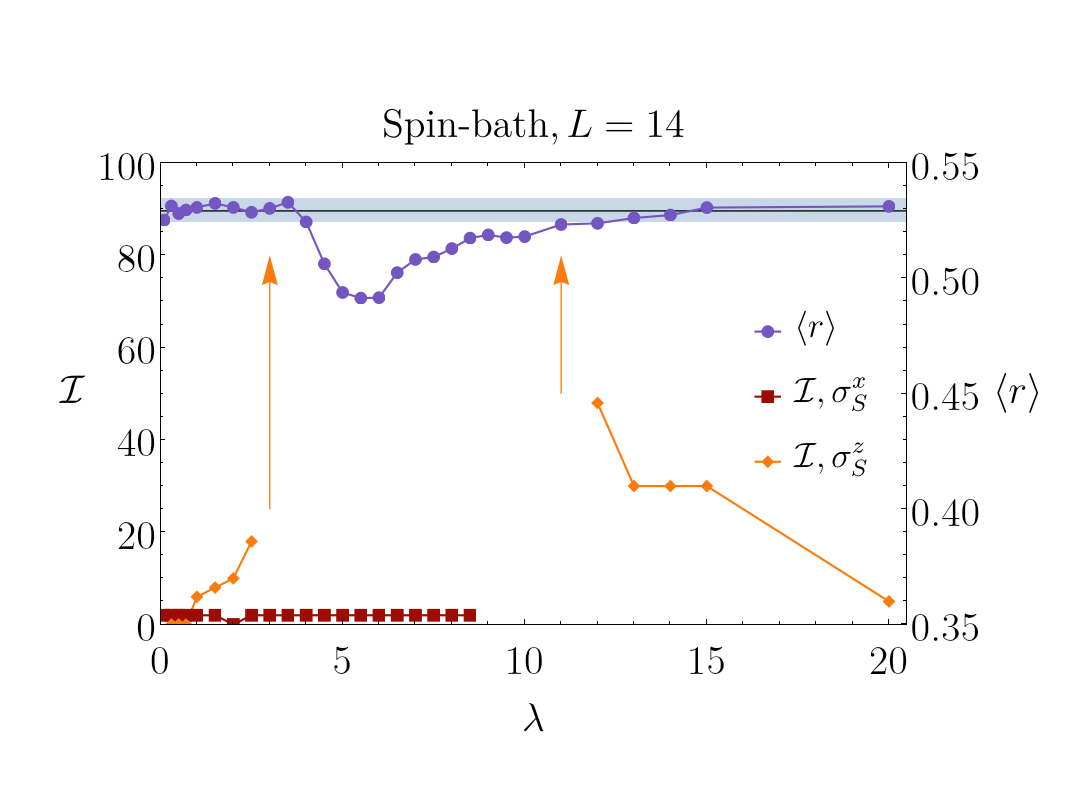}
  \caption{\label{fig:rR_spinbath}AOTF intricacy $\mathcal{I}$ and gap ratio $\langle r\rangle$ for different coupling strengths $\lambda$ in the model of a single spin coupled to an Ising chain with $L=14$. The black line indicates the chaotic benchmark $\langle r\rangle_{\text{GOE}}=0.53$ and the shaded area $1\%$ deviation around it. The arrows indicate that within the parameter range $3\lesssim\lambda\lesssim11$ the intricacies exceed $\mathcal{I}=100$.}
\end{figure}

The spin-bath model (\ref{spinbathhamilton}) exhibits a deviation from full chaoticity  at intermediate spin-bath coupling $4\lesssim\lambda\lesssim11$, cf.\ Fig. \ref{fig:rR_spinbath}. 

At small coupling $\lambda$ the intricacies $\mathcal{I}$ for the AOTFs of $\sigma_S^z$ start off at $\mathcal{I}=0$ as at first the autocorrelation functions are monotonous themselves. With increasing coupling strength, the intricacies grow until $\mathcal{I}=19$ for $\lambda=2.5$. As the system approaches the nonchaotic regime the intricacies exceed $\mathcal{I}=100$, leaving the scope of our numerical method. Once the system approaches the chaotic regime again at $\lambda\sim11$ we again infer intricacies $\mathcal{I}<100$ starting at $\lambda=12$ where $\mathcal{I}=48$, cf.\ Fig.\ \ref{fig:bath_sz}. These then decrease further as the system moves away from the nonchaotic regime, eventually giving rise to an AOTF with $\mathcal{I}=5$ at $\lambda=20$.

In the case of the observable $\sigma_S^x$ the picture is different. The intricacies do not seem to be affected by the systems' transition from the chaotic to an nonchaotic regime. With an exception at $\lambda=2$ where our construction finds the respective autocorrelation function to be suitably monotonous, i.e.\ $\mathcal{I}=0$, we find AOTFs with an intricacy $\mathcal{I}=2$ for any other considered coupling strength $\lambda\le8.5$. Entering a strong coupling range at $\lambda\gtrsim9$ however, the functions $C^2(t)$ no longer decay within a reasonable period in the sense we require to infer an equilibration time $\tau$. Consequently we leave them out of further discussions.

Having presented all numerical evidence, we now aim at distilling the common principle that is  present in all these examples:
Whenever an autocorrelation function of a few-body observable exhibits high intricacy, e.g.   $\mathcal{I}> 20$ (for the $\delta $ chosen here) the corresponding system either belongs to the nonchaotic regime (according to the gap ratio) or a nonchaotic regime is close by with respect to variation of some model parameter.  Note that the converse does not hold: low-intricacy dynamics may emerge in nonchaotic regimes, see e.g.\ $\sigma_S^x$ in the spin-bath model.  This principle is our main numerical result.

The implication of this principle may be loosely phrased as: Deep in the chaotic regime dynamics are simple and time-local arrows of time may be readily constructed.  Intricate dynamics require at least the closeness of a nonchaotic regime.

\section{Arrows of time in the Lanczos picture\label{sec:lanczos}}
In the second part of the paper at hand we wish to put our numerical findings into the context of  well-established concepts such as the recursion method and Lanczos coefficients \cite{viswanath94,parker19}. To this end we briefly introduce pertinent concepts, translate our construction into this setting and expand on relevant quantities.
\subsection{The Recursion Method and Lanczos coefficients}
\begin{figure}[h]
	 \includegraphics[width=0.4\textwidth]{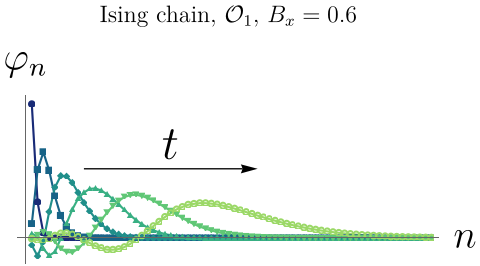}
  \caption{\label{fig:wavevector} Sketch of the wave vector (\ref{entry_wavevector}) for different instances of time $t$ at the example of fast-mode observable $\mathcal{O}_1$ of the energy density in the tilted field Ising chain with $B_x=0.6$.}
\end{figure}

Given some Hamiltonian $\mathcal{H}$ and Hermitian operator $\mathcal{O}$, we are interested in autocorrelation functions
\begin{align*}
    C(t)=\text{tr}\{\mathcal{O}(t)\mathcal{O}\}/d,
\end{align*}
where we denote with $\mathcal{O}(t)$ the Heisenberg time evolution $U(t)^\dagger \mathcal{O}U(t)$ for $U(t)=\exp(-i \mathcal{H}t)$ and $d$ the dimension of the Hilbert space. Working in operator space, also Liouville space, we denote the vectors by $\vert\mathcal{O})$. Equipped with the Hilbert-Schmidt scalar product $(A\vert B):=\text{tr}\{A^\dagger B\}/\text{tr}\{\mathbf{1}\}$ a norm is induced via $\vert\vert\mathcal{O}\vert\vert=\sqrt{(\mathcal{O}\vert\mathcal{O})}$. The Liouville superoperator $\mathcal{L}(-):=\left[\mathcal{H},-\right]$, also Liouvillian, here implements the time evolution as $\vert\mathcal{O}(t))=\exp\left(i\mathcal{L}t\right)\vert\mathcal{O})$. With the Liouvillian it follows that for the autocorrelation we have $C(t)=(\mathcal{O}\vert\exp(i\mathcal{L}t)\vert\mathcal{O})$.
\\
The \textit{Lanczos algorithm} now orthogonalizes the set $\{\mathcal{L}^n\vert\mathcal{O})\}$ as follows. Starting from $\vert\mathcal{O}_0)=:\vert\mathcal{O})$ the algorithm iteratively computes the \textit{Krylov basis} $\{\vert\mathcal{O}_n)\}$ as
 \begin{align}
    \begin{split}
           \vert\tilde{\mathcal{O}}_n)&=\mathcal{L}\vert\mathcal{O}_{n-1})-b_{n-1}\vert\mathcal{O}_{n-2})\\
    b_n&=(\tilde{\mathcal{O}}_n\vert\tilde{\mathcal{O}}_n)^{1/2}\\
    \vert\mathcal{O}_n)&=b_n^{-1}\vert\tilde{\mathcal{O}}_n),\label{lanczosalgorithm}
    \end{split}
\end{align}
where $b_0=0$. \textcolor{black}{The quantities $b_n$ are by construction real and positive and are referred to as \textit{Lanczos coefficients}.} Upon defining the functions
\begin{align}
    \varphi_n(t):=i^{-n}(\mathcal{O}_n\vert\mathcal{O}(t))\label{entry_wavevector}
\end{align} \textcolor{black}{we may highlight the connection of the Lanczos algorithm defined Eq.}\ (\ref{lanczosalgorithm}) \textcolor{black}{and the construction of the AOTFs in Eq.}\ (\ref{construction}). The time derivatives of the autocorrelation function in the AOTF construction \textcolor{black}{are related} to the repeated action of the Liouvillian onto the respective observable. 
Starting from the Heisenberg equation of motion for $\mathcal{O}(t)$ it may be shown (cf. e.g., Refs. \cite{parker19, uskov24}) that the ``wave vector" satisfies a discrete Schr\"odinger equation
\begin{align}
    \label{discreteschroedinger}
    \partial_t \varphi_n(t)=b_n \varphi_{n-1}(t)-b_{n+1}\varphi_{n+1}(t)\quad \text{with}\ n\ge0,
\end{align}
where we set $\varphi_{-1}=0$ by convention and with initial condition $\varphi_n(0)=\delta_{n,0}$. \textcolor{black}{Note that Eq.\ }(\ref{discreteschroedinger}) \textcolor{black}{is equivalent to the construction given in}\ (\ref{construction}). \textcolor{black}{Further technical details on the equivalence between the algorithms described in Eq.\ }(\ref{construction}) \textcolor{black}{and Eqs.\ }(\ref{lanczosalgorithm}),(\ref{entry_wavevector}) \textcolor{black}{are provided in Appendix\ }\ref{app:details}. The functions $\varphi_n(t)$ are equal to the overlap of the $n^{\text{th}}$ Krylov vector with $\mathcal{O}(t)$ as explained in (\ref{entry_wavevector}). For an example  of their (wavelike) dynamics see Fig.\ \ref{fig:wavevector}. \\ For later reference we note that this may be cast into the form of a matrix equation
\begin{align}
    \dot{\boldsymbol{\varphi}}=\textbf{L}.\boldsymbol{\varphi},
\end{align}
where the matrix $\textbf{L}$ is given by
\begin{align}
\label{matrixode}
 \textbf{L} &= \begin{pmatrix}
0 & -b_1 & 0 & 0 & \dots \\
b_1 & 0 & -b_2 & 0 & \dots \\
0 & b_2 & 0 & -b_3 & \dots \\
0 & 0 & b_3 & 0 & \dots \\
\vdots & \vdots & \vdots & \vdots & \ddots \\
\end{pmatrix},
\end{align}
and where $\boldsymbol{\varphi}(t)=(\varphi_0(t),\varphi_1(t),\dots)$.

Renouncing on the ``gauge" (\ref{entry_wavevector}) yields, as an analogue of  (\ref{discreteschroedinger}),
\begin{align*}
    i\partial_t (\mathcal{O}_n\vert\mathcal{O}(t))=-\left[b_n (\mathcal{O}_{n-1} \vert\mathcal{O}(t))+b_{n+1}(\mathcal{O}_{n+1}\vert\mathcal{O}(t))\right].
\end{align*}
This (other) discrete Schr\"odinger equation suggests a frequently adapted interpretation of the Lanczos coefficients as hopping amplitudes of a tight-binding model, cf. Fig \ \ref{fig:sketch_replacement}. The components of the wave function for the $n$-th site on this semi-infinite chain are given by the $ (\mathcal{O}_n\vert\mathcal{O}(t))$.  As $ \varphi_n^2(t) =  |(\mathcal{O}_n\vert\mathcal{O}(t))|^2$ holds, the dynamics of the $\varphi_n(t)$ as generated by  (\ref{discreteschroedinger}) are equivalent to those of a particle hopping on a chain as long as only the properties to sit on specific sites are addressed.
\subsection{Currents and probabilities\label{subsec:currents}} 
The analogy of the tight-binding model gives rise to the interpretation of the entries $\varphi_n(t)$ of the wave vector as amplitudes for a particle being at some site $n$ with probability $ \varphi_n^2(t)$ as we have 
\begin{align}\label{eq_prob_density}
   \sum_{n=0}^{\infty}\varphi_n^2(t)=1.
\end{align}
We then define a \textit{probability operator} $\textbf{P}_k=\mathbf{1}_k\oplus\mathbf{0}$ whose expectation value
\begin{align}
    \langle\textbf{P}_k(t)\rangle=\langle\boldsymbol{\varphi}(t)\vert\textbf{P}_k\vert\boldsymbol{\varphi}(t)\rangle=\sum_{j=0}^k \varphi_j^2(t)\label{probability}
\end{align}
describes the probability to be within the first $k$ sites. From this we can read off that
\begin{align}
    \alpha^\mathcal{I}(t)=\langle\textbf{P}_\mathcal{I}(t)\rangle.\label{aot_prob}
\end{align}
The dynamics of (\ref{probability}) is determined by the Liouvillian $\textbf{L}$ as 
\begin{align}
    \langle\dot{\textbf{P}}_k\rangle=\langle\left[\textbf{P}_k,\textbf{L}\right]\rangle=-2b_k \varphi_{k}(t)\varphi_{k+1}(t)=:-\langle\textbf{J}_k(t)\rangle,\label{loccurrent}
\end{align}
where we define $\textbf{J}_k$ as the associated \textit{probability current} along the $k$-th site.

Hence, an AOTF with intricacy $\mathcal{I}$ is equivalent to the probability of the particle on the semi-infinite chain to be within the first $\mathcal{I}$ sites. This probability is (suitably) monotonously decreasing, i.e.\ there are no (few) \textit{backward} currents of probability.

Moreover, from the comparison of $\alpha^\mathcal{I}$ with the dynamics of the wave vector $\boldsymbol{\varphi}(t)$ we can conclude that whenever $\boldsymbol{\varphi}(t)$ has a zero, i.e.\ two adjacent entries have opposite signs, the AOTF must not be monotonous by (\ref{loccurrent}). Henceforth we argue that by this it is reasonable to allow for small errors $\delta$ from a monotonous behaviour as by continuity is not to be expected that such zeros only occur locally but travel through the chain.

Furthermore the AOTF's are related to the concept of \textit{Krylov-complexity} $K$. With the notation as above the Krylov-complexity reads
\begin{align}
    K(t)=\sum_{n=0}^\infty n\ \boldsymbol{\varphi}_n^2(t)\label{krylovcomplexity},
\end{align}
which in this setting may be interpreted as the expectation value of an observable $\textbf{N}=\text{diag}(0,1,2,\dots)$, \textcolor{black}{i.e.\ $\langle\boldsymbol{\varphi}(t)\vert\textbf{N}\vert\boldsymbol{\varphi}(t)\rangle=K(t)$}. \textcolor{black}{Again invoking the analogy of the Lanczos picture to a particle hopping along a semi-infinite lead as visualized in Fig. \ref{fig:sketch_replacement}, we may also interpret $K(t)$ as the expectation value of the position of the particle. Within this framework the functions $\alpha^n(t) = \langle\textbf{P}_n(t)\rangle $ from (\ref{aot}) and (\ref{probability}), respectively, may be interpreted as the probability to find the  particle at any position  at or left of $n$} \footnote{In fact $\alpha^\mathcal{I}(t)$ itself is a $q$-complexity as $\alpha^\mathcal{I}(t)=\sum_{n=0}^\infty \Theta(\mathcal{I}-n)\boldsymbol{\varphi}^2_n(t)$.}.

\section{structures in the Lanczos coefficients and the arrow of time\label{sec:structures}}
Having translated the AOTF construction put forward in Sec.\ \ref{sec:construction} into the Lanczos picture in Sec.\ \ref{sec:lanczos} we further want to exploit this correspondence by bringing the dynamics of the AOTF into relation with structures in the Lanczos coefficients. To this we motivate an effective replacement model for the dynamics of the wave vector $\boldsymbol{\varphi}(t)$ and discuss its connection to the so-called operator growth hypothesis\ \cite{parker19}.

\subsection{Replacement model}
In order to provide a more intuitive picture of the dynamics of the wave vector $\boldsymbol{\varphi}(t)$ in the Lanczos picture, we aim at mapping Eq. (\ref{discreteschroedinger}) onto a replacement model. We emphasize that the profit of this mapping consists more in the provision of a simple comparable model for many sets of  $b_n$,  rather than its  strict applicability for all possible sets of $b_n$. Discussing the latter in detail would  exceed the scope of the paper at hand. 

To this end we consider the dynamics of a particle with mass $m=1/2$ in one dimension under the influence of a potential $V=-B^2$. With $\psi(x,t)$ as the corresponding wave function the dynamics are given by
\begin{align} \label{repmod}
    i\partial_t \psi(x,t)=\left[-\partial_x^2-B^2(x)\right]\psi(x,t).
\end{align}
Upon a gauge transformation 
\begin{align} \label{gauge}
\psi(x,t)=\exp\left(i\phi(x)\right)\vartheta(x,t),
\end{align}
where $\phi^\prime(x)=B(x)$ one can readily infer that for the dynamics of $\vartheta(x,t)$ we have
\begin{align}
    i\partial_t \vartheta(x,t)=-\left[2iB(x)\partial_x+i B^\prime(x)+\partial_x^2\right]\vartheta(x,t),\label{transformed_ode}
\end{align}
where $B^\prime(x)=\text{d}B(x)/\text{d}x$.  Before proceeding the following should be noted: A lowest order WKB-approximation to the wave function of an outgoing particle with energy eigenvalue $E=0$ for the Hamiltonian in Eq. (\ref{repmod}) reads:
\begin{align}
    \psi_{\text{WKB}}(x) \propto \frac{1}{\sqrt{B(x)}} \exp{ \left(i\int^x B(x') dx'\right)}.
\end{align}
Applying the gauge transformation Eq.\ (\ref{gauge}) to $\psi_{\text{WKB}}$ precisely removes the possibly quickly oscillating phase factor, yielding $\vartheta_{\text{WKB}}\propto 1/\sqrt{B(x)}$. At eigenenergies close to $E=0$ the phase factor will not be removed completely but transformed to oscillate slowly. Thus the following may be expected: For potential landscapes $V=-B^2(x)$ to which the WKB-approximation applies (no turning points of the classical particle, etc.) and for initial states restricted  to low energies, the curvature of $\vartheta (x,t)$ will be and remain low, i.e. $\partial_x^2\vartheta(x,t) \approx 0$ and the corresponding term in Eq.\ (\ref{transformed_ode}) may be neglected.
\begin{figure*}[t]
  \includegraphics[width=\textwidth]{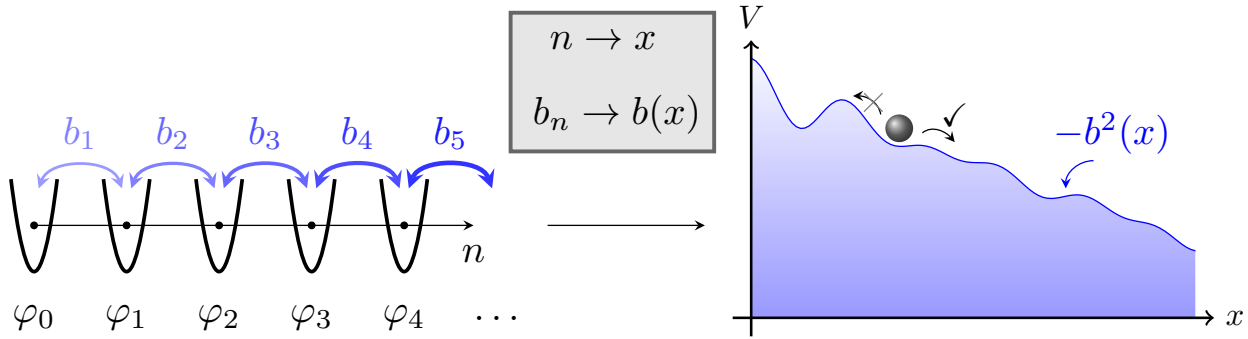}
  \caption{\label{fig:sketch_replacement}  Illustration of the replacement model for the Lanczos dynamics. In the semi-infinite tight-binding model (\textit{left}) the Lanczos coefficients $b_n$ may be viewed as discrete hopping amplitudes. The probability of the particle to be located at some site $j$ is given by $\varphi^2_j$. In the replacement model (see sketch on the \textit{right}) a continuous version of the Lanczos coefficients takes on the role of a potential in the form of $V=-b^2(x)$. This picture motivates the necessity of growing Lanczos coefficients (i.e.\ a decreasing potential) in order for a particle with mass $m=1/2$ to propagate outbound. This connection relates general properties of the respective Lanczos coefficients with the dynamics of the wave vector $\boldsymbol{\varphi}(t)$ and hence to the \textit{arrow of time functions} $\alpha^\mathcal{I}$ via (\ref{aot_prob}).}
\end{figure*}

Turning back to the dynamics of the wave function $\boldsymbol{\varphi}(t)$ in the Lanczos picture, we consider the continuum limit of (\ref{discreteschroedinger}), i.e.\ where $b_n\rightarrow b(x),\ \varphi_{n}(t)\rightarrow \varphi(x,t)$, cf.\ \cite{parker19,zhang24}. Note that whereas the Lanczos coefficient $b_n$ is ``located" between the sites $n-1$ and $n$, the function $b(x)$ takes values at position $x$. Hence $b_n\ \widehat{=}\ b(x-\frac{1}{2})$. 
Up to first order of $b(x)$ and $\varphi(x,t)$ in $x$ we find for (\ref{discreteschroedinger}) that
\begin{align}
    \partial_t \varphi(x,t)=-\left[2b(x)\partial_x+b^\prime (x)\right]\varphi(x,t),\label{lanczos_dirac}
\end{align}
note Ref.\ \footnote{This is a consistent truncation in the sense that $\partial_t \int |\varphi|^2$d$x=0$ strictly holds.}. For more details see Appendix\ \ref{app:linearisation}.
Producing a continuum analog to the initial state indicated below Eq.\ (\ref{discreteschroedinger})  is less unique.  We consider some $\varphi(x,0)$ which is concentrated at $x=0$, i.e., approximately vanishes at $x=1$, but still weakly curved (curvature of order 1).  Associating this state with $\vartheta(x,0)$ and transforming the latter back to $\psi(x,0)$ according to Eq.\ (\ref{gauge}) may yield a low-energy / low-curvature initial state as discussed below Eq.\ (\ref{transformed_ode}) for pertinent $b(x)$. Hence, from comparing (\ref{transformed_ode}) and (\ref{lanczos_dirac}) we infer that the dynamics in the Lanczos picture may be described by those of a particle with mass $m=1/2$ exposed to a potential $V(x)=-b^2(x)$, see also Fig.\ \ref{fig:sketch_replacement}. 

To further strengthen this analogy we consider the dynamics of $\langle\hat{x}\rangle$ in the replacement model. By the Ehrenfest theorem we have for a particle with mass $m=1/2$ that
\begin{align}
    \frac{\text{d}^2\langle \hat{x}\rangle}{\text{d}t^2}=-2\langle\nabla V(\hat{x})\rangle.
\end{align}
Correspondingly, for the respective quantity in the Lanczos picture, i.e. , the Krylov-complexity, we compute
\begin{align}
    \frac{\text{d}^2\langle \textbf{N}\rangle}{\text{d}t^2}&=\langle\left[\left[\textbf{N},\textbf{L}\right],\textbf{L}\right]\rangle=2\sum_j \left(b_{j+1}^2-b_j^2 \right)\varphi_j^2,
\end{align}
with $\textbf{N}$ as defined below Eq.\ \ref{krylovcomplexity}, again suggesting the form of the potential to be $V(x)=-b^2(x)$.  

Having motivated the analogy of the Lanczos dynamics to a particle exposed to a potential $-b^2(x)$ we seek to conclude on properties of $\boldsymbol{\varphi}(t)$ based on the simple replacement model. 

In order for a particle to propagate and to not be trapped, the potential $V$ necessarily needs to fall off, i.e.\ the Lanczos coefficients need to ascend. Qualitatively similar statements may be found in Ref. \cite{parker19}. Furthermore, on qualitative grounds we argue that for a steady ``forward-oriented" motion of the particle, in the sense of no back-transport, the potential additionally needs to exhibit a certain smoothness in order to not give rise to scattering or localization. For this second property we deliberately remain unspecific as we only aim to further stress the suitability of the replacement model rather than to quantify notions of smoothness.

\subsection{Operator Growth Hypothesis}
The operator growth hypothesis (OGH) is a statement on the growth of Lanczos coefficients in infinite-dimensional chaotic many-body systems\ \cite{parker19}. It postulates that in such systems the Lanczos coefficients for local observables asymptotically attain linear growth, i.e.\
\begin{align}
    b_n=\left\{
    \begin{array}{ll}
       A\frac{n}{\ln n}+o\left(\frac{n}{\ln n}\right)& d=1, \\
       \alpha n+\gamma+o(1) & d>1.
    \end{array}
\right.
\end{align}
For numerical purposes however, one is limited to finite systems. In such systems of size $L$, the first $L/2$ Lanczos coefficients $b_n$ coincide with those of the infinite system. Several studies suggest that already the first several Lanczos coefficients feature a linear trend \cite{parker19,dong21,heveling22-2,uskov24,menzler24}. 


\subsection{Lanczos Coefficients and Intricacies}
The analogy between the Lanczos picture and the replacement model predicts that for a steady outward motion of $\boldsymbol{\varphi}(t)$ the Lanczos coefficients need to grow with larger $n$. Moreover the OGH puts forward that in chaotic quantum systems the Lanczos coefficients of local few-body observables eventually attain linear growth. Building up on this connection between Lanczos coefficients and the dynamics of the wave vector $\boldsymbol{\varphi}(t)$ we seek to bridge these structural arguments to the intricacies $\mathcal{I}$ of the respective AOTFs. As pointed out in Sec.\ \ref{subsec:currents}, an AOTF with intricacy $\mathcal{I}$ can be thought of in the Lanczos picture as the (almost) monotonously decreasing probability of the particle to be located within the first $\mathcal{I}$ sites and is thus directly correlated to the dynamics of $\boldsymbol{\varphi}(t)$.

\begin{figure}[]
	 \includegraphics[width=0.5\textwidth]{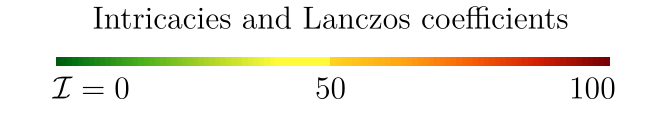}
	 \includegraphics[width=0.45\textwidth]{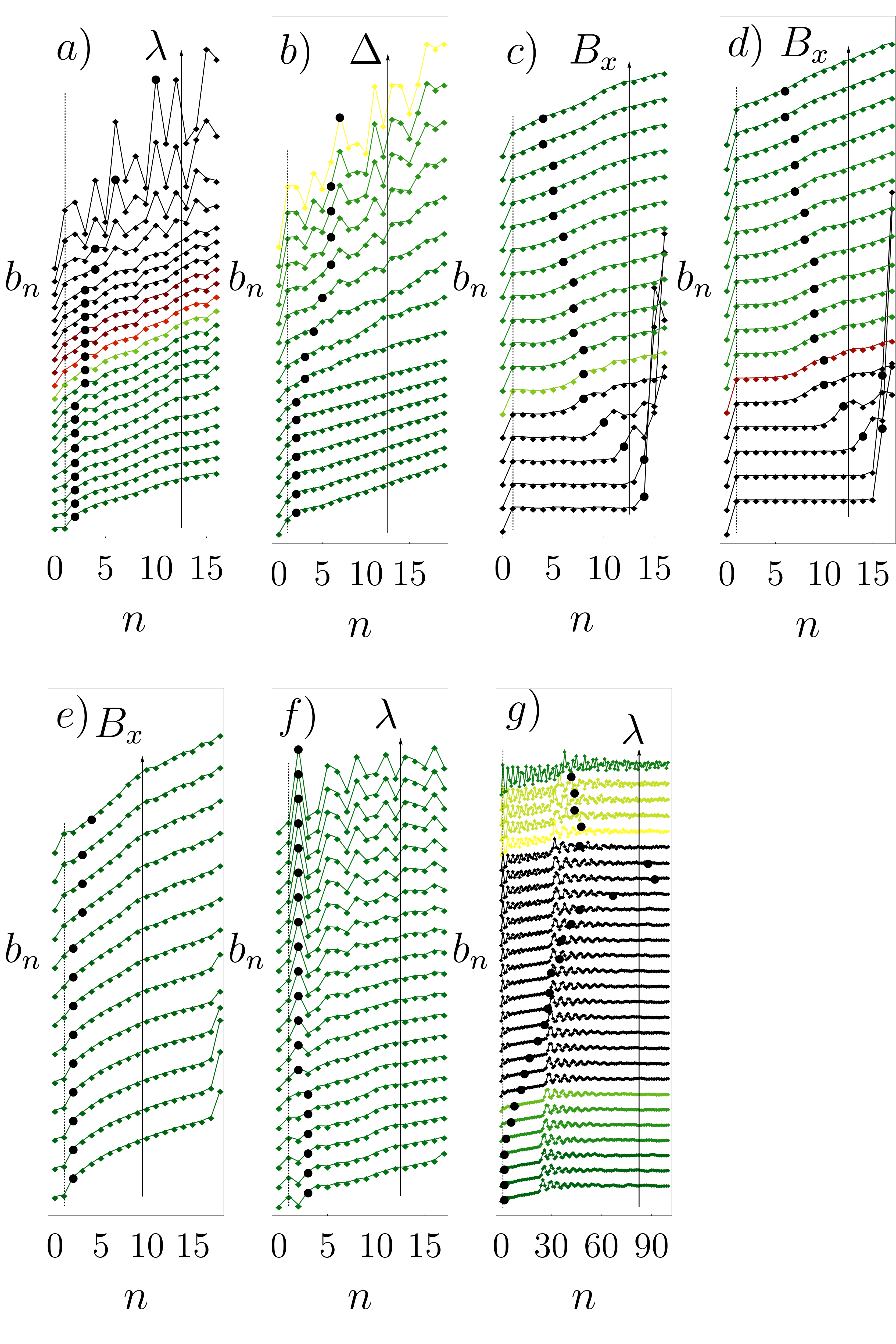}
 \caption{\label{fig:panel_bn} First several Lanczos coefficients for the models and observables considered in Sec.\ \ref{sec:emergence}. In each plot the Lanczos coefficients are shifted vertically for presentation purposes. The dashed line indicates the position of the first Lanczos coefficient. To emphasise the growth, we included $b_0=0$ for each set $\{b_n\}$. The black dots correspond to the points $P$ where the Lanczos coefficients have terminally surpassed $1.5\ b_1$. In cases where $\mathcal{I}>100$ we colour the respective $b_n$ black regardless whether we are able to infer an $\mathcal{I}$ or not. \textit{a}) Ising ladder, \textit{b}) XXZ chain, \textit{c})-\textit{e}) $\mathcal{O}_1-\mathcal{O}_3$ in the tilted field Ising chain respectively, \textit{f}),\textit{g}) $\mathcal{\sigma}_S^x,\mathcal{\sigma}_S^z$ in the spin-bath model. For the coupling strength $\lambda=20$, the Lanczos coefficients for $\sigma_S^z$ do not terminally surpass 1.5\ $b_1$. Therefore we cannot infer $P$ in this case.} 
\end{figure}

Straight-forwardly we determine the position $P$ after which the Lanczos coefficients have terminally surpassed some multiple of $b_1$. Notice that rescaling the set of Lanczos coefficients changes only the time-scale of the associated dynamics. We consider the value $1.5\ b_1$, see Fig.\ \ref{fig:panel_bn}. Note that we consider finite systems. Hence the OGH only pertains to the first $L/2$ Lanczos coefficients, where $L$ is the size of the system. The growth of the Lanczos coefficients does not prevail substantially longer due to the finiteness of the system, e.g.\ cf.\ Fig.\ \ref{fig:panel_bn}\ $g)$ around $n\sim25$. In Fig.\ \ref{fig:panel_bn} we only depict as many Lanczos coefficients as necessary to depict every $P$ in order to avoid cluttering.

We find for each pair $(\mathcal{H},\mathcal{O})$ that whenever the growth of the Lanczos coefficients sets in early (as quantified by $P$), the corresponding intricacy $\mathcal{I}$ is low. The absolute values for $\mathcal{I}$ however are not comparable between different systems and observables. In the derivation for the replacement dynamics the structure of the Lanczos coefficients only enters up to first order. For the two-chain model (\ref{ladder}), the XXZ chain (\ref{xxz-hamiltonian}) as well as for the tilted field Ising model (\ref{isingham}) the Lanczos coefficients mostly do not exhibit strong non-linear behaviour, see Fig.\ \ref{fig:panel_bn} $a)$-$e)$. In these cases $P$ and the intricacies $\mathcal{I}$ correlate. This observation also holds true for the spin-bath model (\ref{spinbathhamilton}) for small coupling $\lambda$, as depicted in Fig.\ \ref{fig:panel_bn} $f)$, $g).$ However starting at intermediate coupling strength $\lambda\sim 3$ the Lanczos coefficients exhibit a non-linear, staggered structure, already for the infinite-system $b_n$. Whereas at first the correlation between $P$ and $\mathcal{I}$ is visible for small spin-bath coupling strengths $\lambda$ for $\mathcal{O}=\sigma_S^z$, the analogy fails in the strong-coupling regime. At $\lambda=20$ the intricacy becomes $\mathcal{I}=5$, although $P$ is largest for this interaction strength, exceeding the range of Lanczos coefficients considered for this example. However, as already pointed out above, the dynamics for the replacement model only takes the structure of the $b_n$ up to first order into account.

Nevertheless, we find that within its range of applicability the reasoning based on the replacement model strongly hints on a connection between the intricacies $\mathcal{I}$ and growth of the respective Lanczos coefficients. Furthermore we expect that whenever the OGH holds true in the sense that the linear growth not only prevails asymptotically but also sets in conceivably early (as observed in several studies), the intricacies of AOTFs are small.

\section{conclusion and remarks\label{sec:conclusion}}
In this work we proposed a construction for monotonously decreasing arrow of time functions (AOTFs) for autocorrelation functions of local observables. The construction of the AOTFs requires $2\mathcal{I}$ temporal derivatives of the respective autocorrelation function itself. Due to this, and our qualitative assessment that autocorrelation function yielding larger $\mathcal{I}$ feature  richer structures, we refer to  $\mathcal{I}$ as \textit{intricacy}.

We numerically investigated the AOTFs  of several few-body observables in pertinent quantum models. 
In these we found that whenever the system is fully chaotic (subject to the gap ratio) the intricacies $\mathcal{I}$ are low. Conversely, only when the system is nonchaotic or close to a nonchaotic regime (with respect to some model parameter) the dynamics may become highly intricate. However, integrability itself (or closeness to it) is no sufficient condition for high-intricacy dynamics.

Furthermore we translated our findings into the setting of Lanczos coefficients and the recursion method. 
In this picture the AOTF is equivalent to the probability of a particle to be located within the first $\mathcal{I}$ sites in a tight-binding model, where the Lanczos coefficients $b_n$ take the role of the hopping amplitudes.

By linearising the continuum limit of the equations of motion in the Lanczos picture we introduced a replacement model providing an analogy to the motion of particle in a potential given by the negative square of the Lanczos coefficients $b_n$. In this analogy the motion of the particle in the replacement model represents the flux of probability in the tight-binding model.  Consequently if the motion in the replacement model is predominantly outbound, there are only few backward currents of probability in the Lanczos picture which in turn gives rise to an arrow of time. In order for this to occur the potential needs to decrease, which in turn requires growing Lanczos coefficients. This is to be compared to the operator growth hypothesis which predicts growing Lanczos coefficients for chaotic systems.
We find in most of the cases the linearisation to be appropriate for the first several Lanczos coefficients. In these cases the growth of the $b_n$ and the intricacies correlate, linking our our numerical findings to the operator growth hypothesis.
\section*{Acknowledgements}
We thank Christian Bartsch, Robin Heveling, Mats H. Lamann, Alvaro M. Alhambra, Robin Steinigeweg and Philipp Strasberg for fruitful discussions. J.W. thanks Heiko G. Menzler for discussions on the numerical method.
This work has been funded by the Deutsche Forschungsgemeinschaft (DFG), under Grant No. 531128043, as well as under Grant No. 397107022, No. 397067869, and No. 397082825 within the DFG Research Unit FOR 2692, under Grant No. 355031190.
\vspace{0.3cm}
\appendix
\section{\textcolor{black}{LANCZOS ALGORITHM WITH TEMPORAL DERIVATIVES OF THE AUTOCORRELATION FUNCTION}\label{app:details}}
The construction of the AOTFs in Eq.\ (\ref{construction}) may be understood as a reformulation of the standard Lanczos scheme Eq.\ (\ref{lanczosalgorithm}) in the basis given by Eq.\ (\ref{entry_wavevector}). Rewriting the standard Lanczos step as
\begin{align}
   \vert\tilde{\mathcal{O}}_n)= b_n\vert\mathcal{O}_n)=\mathcal{L}\vert\mathcal{O}_{n-1})-b_{n-1}\vert\mathcal{O}_{n-2}),\label{app_eq_lanczos}
\end{align}
we apply $(\mathcal{O}(t)\vert$ and conjugate to find
\begin{align}
    b_n(\mathcal{O}_n\vert\mathcal{O}(t))&=(\mathcal{O}_{n-1}\vert\mathcal{L}\vert\mathcal{O}(t))-b_{n-1}(\mathcal{O}_{n-2}\vert\mathcal{O}(t)).\end{align}
    Using the Heisenberg equation
\begin{align}
    \mathcal{L}\vert\mathcal{O}(t))=\vert\left[\mathcal{H},\mathcal{O}(t)\right])=-i\partial_t\vert \mathcal{O}(t)),
\end{align} and the gauge (\ref{entry_wavevector}) we arrive at
\begin{align}
    b_n i^n\varphi_n(t)&=-i^n\partial_t\varphi_{n-1}(t)-b_{n-1}i^{n-2}\varphi_{n-2}\\\Leftrightarrow b_n\varphi_n(t)&=-\partial_t\varphi_{n-1}(t)+b_{n-1}\varphi_{n-2}(t)\label{eq_app_schroedinger}\\ &=\tilde{\varphi}_n(t)
\end{align}
which is the first line in the construction scheme given in Eq.\ (\ref{construction}). Note that the penultimate line is the discrete Schr\"odinger equation given in Eq.\ (\ref{discreteschroedinger}). To argue that an appropriate scalar product is given by
\begin{align}
    \langle\partial_t^n C(t),\partial_t^m C(t)\rangle:=(-1)^n \partial_t^{n+m}C(t)\vert_{t=0}
\end{align} several remarks are in order. First note that whenever the integers $n,m$ have different parity the scalar product vanishes because an odd moment is addressed. Moreover as $n,m$ need to have the same parity to yield a nonzero scalar product the above scalar product is symmetric as $(-1)^n=(-1)^m$. Consider the following:
\begin{align}
    C(t)&=(\mathcal{O}\vert e^{i\mathcal{L}t}\vert\mathcal{O})\\\rightarrow \partial_t^nC(t)&=i^n(\mathcal{O}\vert \mathcal{L}^n e^{i\mathcal{L}t}\vert\mathcal{O})\\
    \rightarrow \partial_t^n C(t)\vert_{t=0}&=i^n(\mathcal{O}\vert\mathcal{L}^n\vert\mathcal{O})\label{app-eq-c0}.
\end{align}
\textcolor{black}{From}\ (\ref{app-eq-c0}) \textcolor{black}{we may infer that the above scalar product is positive definite as}
\begin{align}
\begin{split}
        \langle \partial_t^n C(t),\partial_t^n C(t)\rangle&=(-1)^n \partial_t^{2n}C(t)\vert_{t=0}\\ &=(\mathcal{O}\vert\mathcal{L}^{2n}\vert\mathcal{O})\\
        &=\vert\vert\mathcal{L}^n\vert\mathcal{O})\vert\vert\ge0,
\end{split}
\end{align}
\textcolor{black}{where we used that $(-1)^n=(-1)^{-n}$ for any natural $n$.}

We now show that the Lanczos coefficients obtained from $\vert\tilde{\mathcal{O}}_n)$ and $\tilde{\varphi}_n(t)$ (as appearing in (\ref{lanczosalgorithm}) and (\ref{construction}) respectively) coincide, implying the appropriateness of the above scalar product. To this end, consider $\vert\tilde{\mathcal{O}}_n)$ expanded in the basis $\{\mathcal{L}^k\vert\mathcal{O})\}_{k=0}^n$
\begin{align}
    \vert\tilde{\mathcal{O}}_{n})=\sum_{k=0}^n c_k\mathcal{L}^k\vert\mathcal{O}),
\end{align}
where the $c_k$ are real numbers determined by the Lanczos coefficients.
From the structure of the Lanczos algorithm all $c_k$ where $k$ has a different parity then $n$ are zero. For the respective Lanczos coefficient we may write
\begin{align}
\begin{split}
        b_n^2=(\tilde{\mathcal{O}}_n\vert\tilde{\mathcal{O}}_n)&=\sum_{k,j}c_j c_k(\mathcal{O}\vert\mathcal{L}^{j+k}\vert\mathcal{O})\\
    &=\sum_{k,j}c_j c_k i^{-(k+j)}\partial_t^{k+j}C(t)\vert_{t=0},\label{app-bn-on}
\end{split}
\end{align}
where we made use of Eq.\ (\ref{app-eq-c0}).
In order to get the respective representation of $\tilde{\varphi}_n(t)$ consider
\begin{align}
\begin{split}
        (\tilde{\mathcal{O}}_n\vert\mathcal{O}(t))&=\sum_k c_k(\mathcal{O}\vert\mathcal{L}^k\vert\mathcal{O}(t))=\sum_k c_k(\mathcal{O}\vert\mathcal{L}^k e^{i\mathcal{L}t}\vert\mathcal{O})\\
    &=\sum_k c_k i^{-k}\partial_t^k C(t).
\end{split}
\end{align}
Via the gauge (\ref{entry_wavevector}) both pictures are related. Concretely
\begin{align}
\begin{split}
       \tilde{\varphi}_n(t)= b_n\varphi_n(t)&=i^{-n}(\tilde{\mathcal{O}}_n\vert\mathcal{O}(t))\\
   &=i^{-n}\sum_k c_k i^{-k}\partial_t^k C(t).
\end{split}
\end{align}
Note that the $\tilde{\varphi}_n(t)$ are real because since $n$ and $k$ need to have the same parity the factors $i^{-(n+k)}$ are never imaginary.
For the respective Lanczos coefficient we may write
\begin{align}
\begin{split}
      \langle\tilde{\varphi}_n(t),\tilde{\varphi}_n(t)\rangle&=(-1)^n\sum_{k,j}c_k c_j i^{-(k+j)}\langle\partial_t^j C(t),\partial_t^k C(t)\rangle\\
    &=(-1)^n\sum_{k,j}c_k c_j i^{-(k+j)}(-1)^k\partial_t^{j+k}C(t)\vert_{t=0}\\&=\sum_{k,j}c_k c_j \underbrace{(-1)^{n+k}}_{+1} i^{-(k+j)}\partial_t^{j+k}C(t)\vert_{t=0}.\label{app-bn-phi}
\end{split}
\end{align}
Again noting that $j,k$ need to have the same parity as $n$, eventually we may infer that Eq.\ (\ref{app-bn-on}) and Eq.\ (\ref{app-bn-phi}) coincide.
\section{NUMERICAL METHOD\label{numerical_method}}
For the numerics in Sec.\ \ref{sec:emergence} we determine the functions $\varphi_j(t)$ defined in (\ref{lanczosalgorithm}) as entries of the wave vector $\boldsymbol{\varphi}$ via (\ref{entry_wavevector}). To ensure orthogonality between the vectors $\vert\mathcal{O}_n)$ in the Krylov basis we perform \textit{partial re-orthogonalisation} \cite{simon84,rabinovici21}. Additionally we numerically verify that $\max_n\vert\partial_t \varphi_n-b_n \varphi_{n-1}+b_{n+1}\varphi_{n+1}\vert\le10^{-8}$, indicating the accuracy of our result.
\section{DERIVATION OF (\ref{lanczos_dirac})\label{app:linearisation}}
Linearising both $\varphi(x)$ and $b(x)$ in $x$ we have
\begin{align*}
    \varphi^{}_{n\pm1}(t)&\rightarrow \varphi(x)\pm\varphi^\prime_{}(x),\\
    b_{n}&\rightarrow b(x)-\frac{1}{2}b^\prime(x),\\
    b_{n+1}&\rightarrow b(x)+\frac{1}{2}b^\prime(x).
\end{align*}
With this we one can rewrite (\ref{discreteschroedinger}) to find (\ref{lanczos_dirac}).
\bibliography{main}
\end{document}